\documentclass[twocolumn,showpacs,preprintnumbers,amsmath,amssymb,prb]{revtex4}

\usepackage{graphicx}
\usepackage{dcolumn}
\usepackage{bm}

\begin{document}

\title{
Solution of electric-field-driven tight-binding lattice coupled to
fermion reservoirs
}

\author{Jong E. Han}
\affiliation{
Department of Physics, State University of New York at Buffalo, Buffalo,
New York 14260, USA}

\date{\today}

\begin{abstract} 
We study electrons in tight-binding lattice driven by DC electric field
with their energy dissipated through on-site fermionic thermostats. Due to the
translational invariance in the transport direction, the problem can be
block-diagonalized. We solve this time-dependent quadratic problem and
demonstrate that the problem has well-defined steady-state. The
steady-state occupation number shows that the Fermi surface shifts at
small field by the drift velocity, in agreement with the Boltzmann
transport theory, but it then deviates significantly at high
fields due to strong nonlinear effect. Despite the
lack of momentum scattering, the conductivity takes the same form as the
semi-classical Ohmic expression from the relaxation-time approximation.  
\end{abstract}


\maketitle

\section{Introduction}

Nonequilibrium phenomena in lattice are the oldest and most
fundamental problems in solid state physics. In conventional solids,
acceleration due to external field is relatively small compared to
electronic energy, and various scattering mechanisms make the transport
diffusive enough so that the small field approximation has often been
applicable. The quantum Boltzmann method has been applied
effectively~\cite{kadanoff,mahan} and linear response limit has been
widely used in the solid-state literature. However, recent progress in nano-devices
and optical lattice systems has made rigorous high-field formalism necessary to
understand their non-perturbative effects such as the Bloch oscillation.
In such regime, understanding the interplay of non-perturbative
field-effect and the many-body physics has emerged as one of the most
pressing problems in nano-science.

Combining the nonequilibrium and quantum many-body effects is an
extremely challenging task. Much effort has been exerted towards
understanding strong correlation physics in quantum dot physics,
especially the prototypical nonequilibrium Kondo problem. Analytical theories
\cite{rosch,schoeller,mehta} and many numerical methods have been
proposed along the time-dependent~\cite{werner,feiguin,schiro}, and
steady-state simulations~\cite{prl07,anders}. In such systems with
localized interacting region, the important question of energy
dissipation could have been side-stepped, and the existence of
steady-state has not been a major issue.

In the past few years, non-perturbative inclusion of electric-field and
many-body effects in lattice systems has been one of the central issues
in the field.  Theories for lattice nonequilibrium have been
formulated~\cite{turkowski,freericks}, mostly based on the dynamical
mean-field theory (DMFT) for an $s$-orbital tight-binding (TB) lattice
with on-site interaction~\cite{eckstein,aoki,amaricci,aron}. Various
attempts have been made to include dissipation mechanism to the driven
lattice by fermion bath~\cite{aoki} and bosonic baths~\cite{vidmar}.
This work corresponds to the analytic solution of the non-interacting
limit of the models considered in Refs.~\onlinecite{aoki,amaricci}.
Although a long-held belief in solid-state transport has been that,
under a finite electric-field, the Fermi sea is perturbatively shifted
by drift velocity, many calculations performed under the DMFT framework
have suggested that the system approaches a steady-state with infinitely
hot electron gas even for small field. With inclusion of proper
dissipation mechanism, one expects the Boltzmann picture of displaced
Fermi surface at small fields and a recovery of the Bloch oscillation in
the high-field limit.

However, it has been unclear so far what approximations, such as
single-band approximation without Landau-Zener tunneling or the nature
of on-site interaction, are responsible for the rather peculiar
long-time states obtained from numerical theories. One of the goals of
this paper is that we provide exact solutions to one of the simplest
dissipation models with on-site fermion thermostats and give analytic
understanding of the problem, and guide possible future modeling.

Due to the nature of the one-body reservoirs, the problem can be solved
exactly (see Fig.~\ref{fig1}). With identical reservoirs on each site,
the Hamiltonian can be block-diagonalized according to the wave-vector
of electrons in the transport direction. The block-diagonal Hamiltonian
can then be exactly solved by a time-dependent perturbation
theory~\cite{jauho,blandin} using the nonequilibrium Green function
theory. The calculation of the wave-vector dependent occupation number
supports the semi-classical Boltzmann transport theory despite the
lack of momentum scattering.
DC electric current of this model is shown analytically
to recover the familiar semi-classical Boltzmann equation
result~\cite{lebwohl}. Based on these findings, we conclude that the
fermion thermostat model, despite its crude modeling to realistic
dissipation mechanism, can serve as a minimal setup for the studies of
strong correlation effects in driven lattice models. Although the model
considered here is one-dimensional, the result can be readily extended
to any spatial dimensions since the model is one-body and conserves
momentum.

\section{Model}
\begin{figure}
\includegraphics[width=\linewidth]{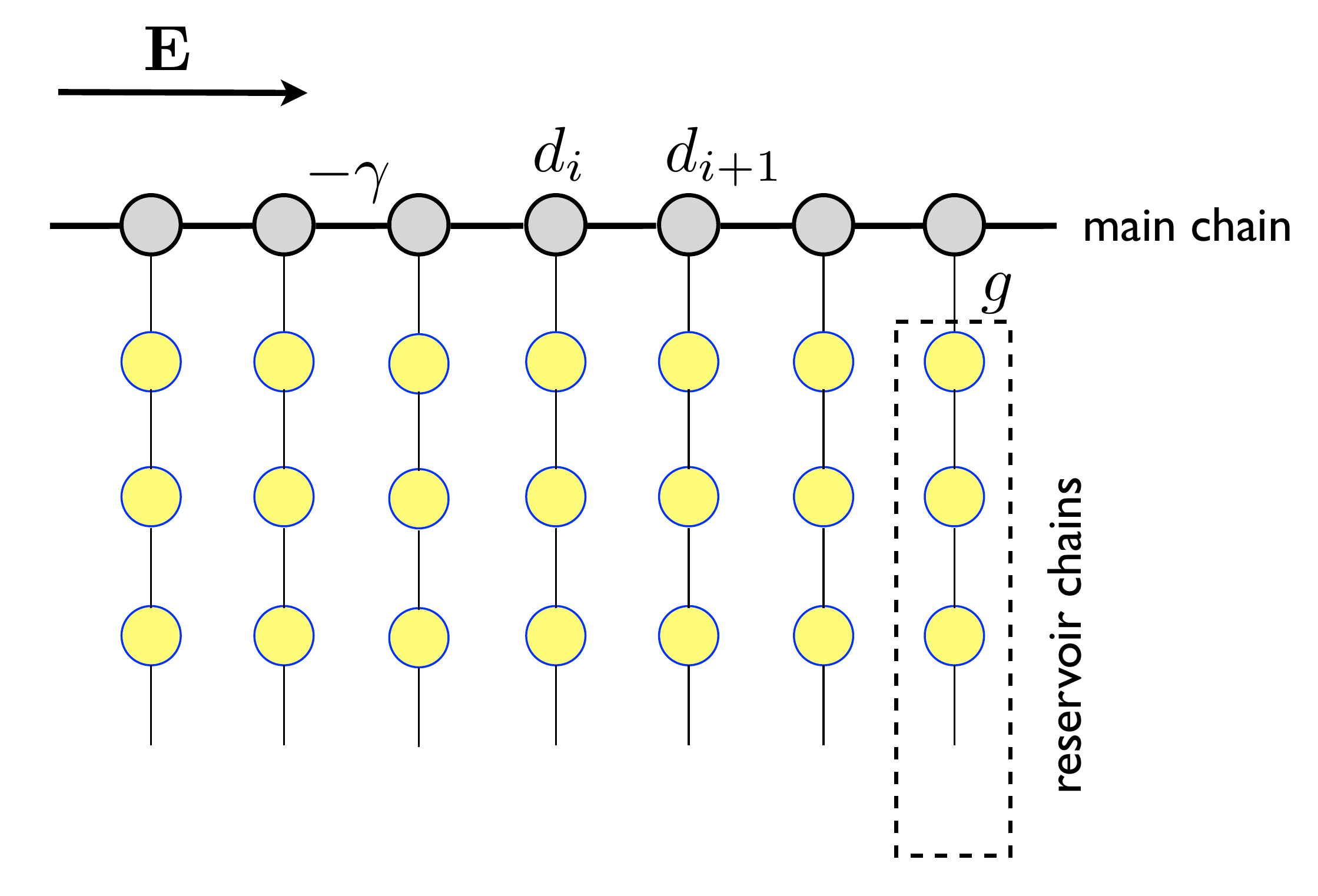}
\caption{(Color online) One-dimensional tight-binding lattice of orbital $d_i$ under an
electric field $E$. Each lattice site is connected to an identical
fermionic bath of $\{c_{i\alpha}\}$ with the continuum index $\alpha$
along the reservoir chain direction.
}
\label{fig1}
\end{figure}

We study a quadratic model of a one-dimensional $s$-orbital
tight-binding model connected to fermionic reservoirs (see
Fig.~\ref{fig1}) under a uniform electric field $E$. The effect of the electric
field is absorbed in the temporal gauge as the Peierls phase
$\varphi(t)=(eEa)t$ to the
hopping integral~\cite{turkowski} $\gamma$.
The time-dependent Hamiltonian then reads
\begin{eqnarray}
\hat{H}(t)&=&-\gamma\sum_i (e^{i\varphi(t)}
d^\dagger_{i+1}d_i+H.c.)+\sum_{i\alpha}\epsilon_\alpha
c^\dagger_{i\alpha}c_{i\alpha}\nonumber \\
&  &-g\sum_{i\alpha}(c^\dagger_{i\alpha}d_i+H.c.),
\end{eqnarray}
with $d^\dagger_i$ as the (spinless) electron operator on the
tight-binding chain on site $i$, $c^\dagger_{i\alpha}$ with the
reservoir fermion states connected to the site $i$ with the continuum index
$\alpha$ along each reservoir chain. Here we do not specify the explicit connectivity of the reservoir
chains, but each chains are assumed to have an identical dispersion
relation $\epsilon_\alpha$.  Notice that the electric field is applied
only on the tight-binding chain $\{d^\dagger_i\}$. The coupling between
the TB site and the reservoir is given by the identical tunneling
parameter $g$. The Peierls phase $\varphi(t)$ is
given as
\begin{equation}
\varphi(t) = \left\{\begin{array}{cc}
0, & \mbox{for }t<0 \\
\Omega t, & \mbox{for }t>0 
\end{array}\right.
\end{equation}
$\Omega=eEa$ is the
Bloch-oscillation frequency due to the electric field.

We note that the whole system has discrete translational symmetry in the
transport direction and
the Hamiltonian is readily block-diagonalized with respect to the
wave-vector $k$ as $d^\dagger_k=\sqrt{N_d^{-1}}\sum_j
e^{ikR_j}d^\dagger_j$ and $c^\dagger_{k\alpha}=\sqrt{N_d^{-1}}\sum_j
e^{ikR_j}c^\dagger_{j\alpha}$ (with lattice sites $R_j=aj$ and the number of
sites along the TB chain $N_d$),
\begin{eqnarray}
\hat{H}(t)&=&\sum_k\left[
-2\gamma\cos(k+\varphi(t))d^\dagger_kd_k+\sum_{\alpha}\epsilon_\alpha
c^\dagger_{k\alpha}c_{k\alpha}\right.\nonumber \\
&& \left. -g\sum_{\alpha}(c^\dagger_{k\alpha}d_k+H.c.)\right].
\end{eqnarray}
Here $\epsilon_{d}(k)=-2\gamma\cos(k)$ is the tight-binding dispersion
at zero $E$-field.
Then each $k$-sector can be treated and solved separately. So from now
on, we suppress the $k$-subscript until necessary with the following Hamiltonian,
\begin{eqnarray}
\hat{H}_k(t)&=&
-2\gamma\cos(k+\varphi(t))d^\dagger d+\sum_{\alpha}\epsilon_\alpha
c^\dagger_{\alpha}c_{\alpha}\nonumber \\
&& -g\sum_{\alpha}(c^\dagger_{\alpha}d+H.c.).
\label{eq:hk}
\end{eqnarray}
It is important to note that the $k$-dependence enters the problem as
$k+\varphi(t)$ for $t>0$. This problem is simply a resonant level
model~\cite{jauho}
where the level is modulated sinusoidally for $t>0$. 

\section{Solution for occupation number and current}
The time-dependent Hamiltonian~(\ref{eq:hk}) can be exactly solved by
the nonequilibrium Keldysh Green function method~\cite{blandin}. We
write the Hamiltonian as $\hat{H}_k(t)=\hat{H}_0+\hat{V}(t)$ with the
time-independent unperturbed part $\hat{H}_0=\hat{H}_k(0)$ and the
time-dependent perturbation as $\hat{V}(t)=\hat{H}_k(t)-\hat{H}_k(0)$,
\begin{equation}
\hat{V}(t) = -2\gamma\left[\cos(k+\varphi(t))-\cos(k)\right]d^\dagger d
\equiv v(t)d^\dagger d.
\end{equation}
When the perturbation is one-body on discrete states the lesser and
greater part of the self-energy is zero, and the lesser $d$-Green
function $G^<$ is expressed only in terms of the transient term,
symbolically written in the matrix form as~\cite{blandin}
\begin{equation}
{\bf G}^<=[I+{\bf G}^r{\bf V}]{\bf G}_0^<[I+{\bf V}{\bf G}^a]\label{eq:glss}
\end{equation}
and the retarded Green function ${\bf G}^r$ is given by the usual Dyson's
equation
\begin{equation}
{\bf G}^r={\bf G}_0^r+{\bf G}_0^r{\bf V}{\bf G}^r,\label{eq:gr}
\end{equation}
where the matrix product denotes convolution-integrals in time. 

First with the retarded functions, the non-interacting limit has the
time-translational symmetry and
\begin{eqnarray}
G_0^r(t-t')&=&-i\theta(t-t')\int_{-\infty}^\infty d\epsilon
\frac{\Gamma/\pi}{\epsilon^2+\Gamma^2}e^{-i\epsilon(t-t')}\nonumber \\
&=&-i\theta(t-t')e^{-i\epsilon_d(k)(t-t')-\Gamma|t-t'|},
\end{eqnarray}
where we we use a flat-band DOS for the
reservoir in the infinite-band limit with the hybridization broadening
$\Gamma=\pi g^2 N(0)$ and the density of states of the fermion bath
$N(0)=\sum_\alpha\delta(\epsilon_\alpha)$. 
Writing $G^r(t,t')=G_0^r(t-t')g^r(t,t')$,
Eq.~(\ref{eq:gr}) becomes
\begin{equation}
g^r(t,t')=1-i\int_{t'}^t ds\,v(s)g^r(s,t'),
\end{equation}
which can be solved as
\begin{equation}
g^r(t,t')=\exp\left[-i\int_{t'}^t v(s)ds\right],
\end{equation}
and finally we have for the full retarded Green function
\begin{equation}
G^r(t,t')=-i\theta(t-t')e^{-i\epsilon_d(k)(t-t')-\Gamma|t-t'|}
\exp\left[-i\int_{t'}^t v(s)ds\right].
\end{equation}

\begin{figure}
\includegraphics[width=\linewidth]{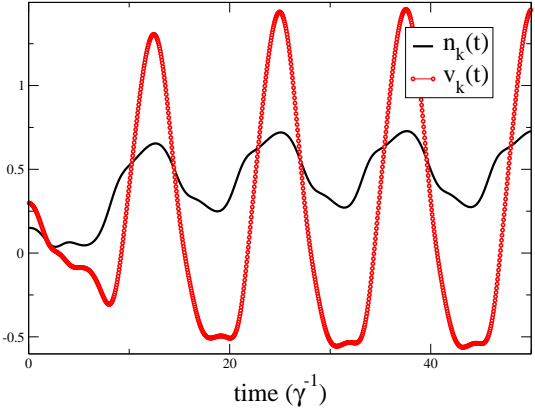}
\caption{(Color online) Expectation value of occupation number $n_k(t)$ of wave-vector
$k$ and the current $J_k(t)$ for $\gamma=1$, $\Omega=0.5$, $\Gamma=0.1$
and at $k=\pi/2+0.1$. After the initial time of $\Gamma^{-1}$, transient
behavior dies out and the expectation values reach steady oscillation.
}
\label{fig:nk}
\end{figure}

For the same-time argument
for $G^<$ we have the Dyson's equation
\begin{eqnarray}
&&G^<(t,t)  = G_0^<(t,t) \nonumber\\
\quad && +\int_0^t \left[G^r(t,s)v(s)G_0^<(s,t)+
 G_0^<(t,s)v(s)G^a(s,t)\right]ds\nonumber\\
\quad && +\int_0^t \int_0^t 
G^r(t,s)v(s)G_0^<(s,s')v(s')G^a(s',t)dsds'.
\end{eqnarray}
We set the initial lesser Green function with the half-filled reservoir
at zero temperature as
\begin{equation}
G_0^<(t,t')=i\int_{-\infty}^0 d\omega
\frac{\Gamma/\pi}{(\omega-\epsilon_d(k))^2+\Gamma^2}e^{-i\omega(t-t')}.
\end{equation}
After some straightforward steps, the occupation number for the 
wave-vector $k$, $n_k(t)=-iG^<(t,t)$, becomes
\begin{eqnarray}
n_k(t) & = & \int_{-\infty}^0 d\omega
\frac{\Gamma/\pi}{(\omega-\epsilon_d(k))^2+\Gamma^2}\times\\
& &\left|1-i\int_0^t ds\,v(s)e^{i(\omega-\epsilon_d(k)+i\Gamma)(t-s)
-i\int_{s}^t v(s')ds'}
\right|^2.\nonumber
\end{eqnarray}
Fig.~\ref{fig:nk} shows the above $n_k(t)$ numerically evaluated for
$\gamma=1$, $\Omega=0.5$, $\Gamma=0.1$ and at $k=\pi/2+0.1$.
Due to the exponential factor $e^{-\Gamma(t-s)}$, the integral converges
to a steady-state oscillation state after time $t\approx \Gamma^{-1}$ and the
transient behavior dies out.
Therefore, for long-time behavior, the time-integral range $[0,t]$ can be changed
to $[-\infty,t]$ for easier analytic treatment. After an
integral-by-parts and some straightforward steps, we have
\begin{eqnarray}
n_k(t) & &= \frac{\Gamma}{\pi}\int_{-\infty}^0 d\omega
\times
\label{eq:nkt}\\
& &\left|\int_{-\infty}^0 ds\,e^{-i(\omega+i\Gamma)s
-i(2\gamma/\Omega)\sin(k+\Omega (t+s))}
\right|^2.\nonumber
\end{eqnarray}
An identity for Bessel functions $J_n(x)$
\begin{equation}
e^{ix\cos\theta}=\sum_{n=-\infty}^\infty i^nJ_n(x)e^{in\theta}
\end{equation}
can be used to perform the integrals as
\begin{eqnarray}
n_k(t) & = &
\frac{\Gamma}{\pi}\sum_{nm}\frac{J_n(\frac{2\gamma}{\Omega})J_m(\frac{2\gamma}{\Omega})e^{i(m-n)(k+\Omega
t)}}{-(m-n)\Omega+2i\Gamma}\times\nonumber\\
&&\left[\frac12\log\frac{m^2\Omega^2+\Gamma^2}{n^2\Omega^2+\Gamma^2}
+i\chi_{mn}\right]
\label{eq:nk}
\end{eqnarray}
with
\begin{equation}
\chi_{mn}=
\pi+\tan^{-1}\frac{m\Omega}{\Gamma}+
\tan^{-1}\frac{n\Omega}{\Gamma}.
\end{equation}

\begin{figure}
\includegraphics[width=0.85\linewidth]{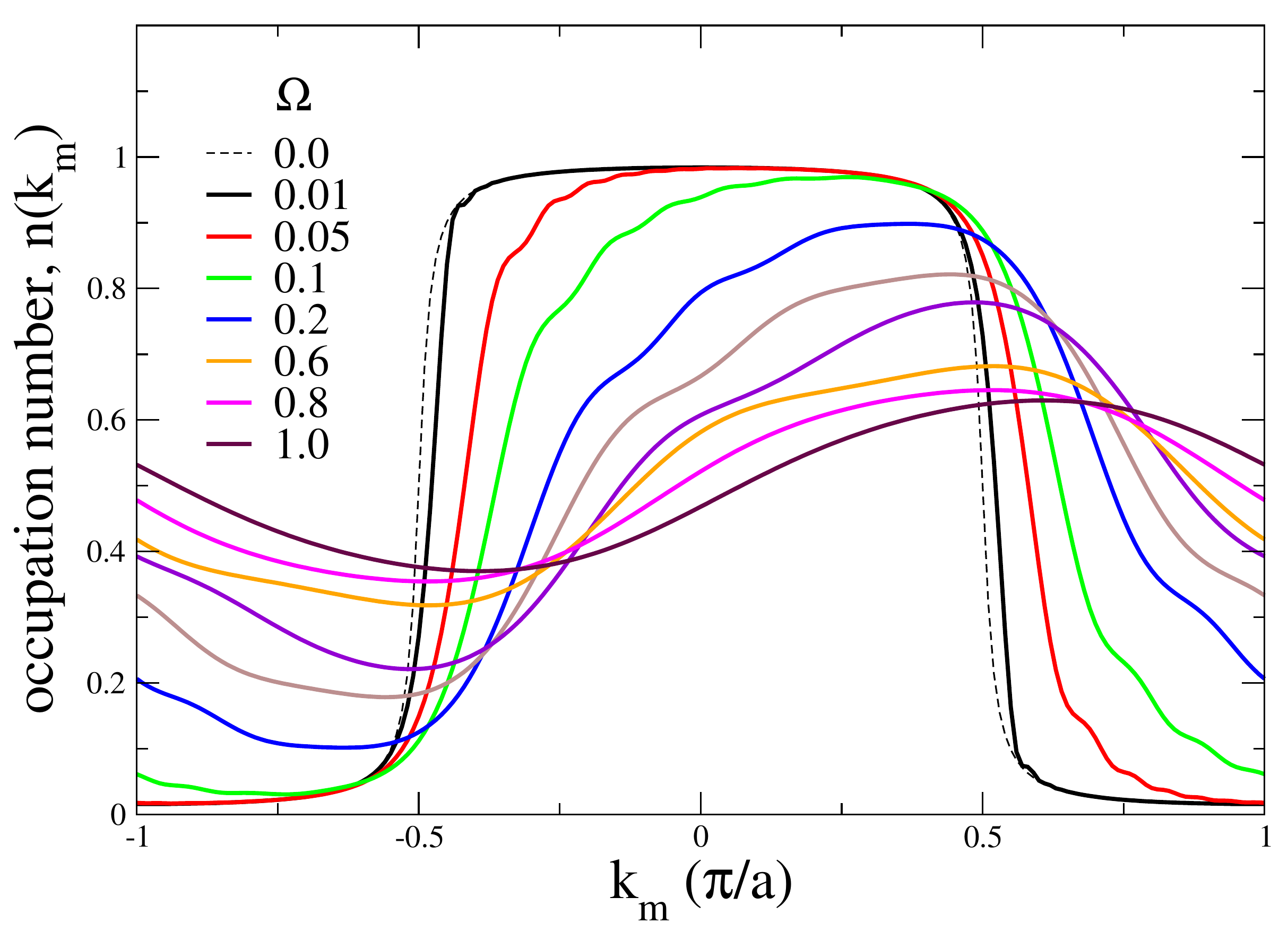}
\caption{(color online) Occupation number $n(k_m)$ with respect to the gauge-invariant
mechanical wave-vector $k_m=k+\Omega t$ from Eq.~(\ref{eq:nk}) at
$\Gamma=0.1$. At zero
field ($\Omega=0$, dashed line), $n(k_m)$ is given by the Fermi-dirac
distribution with the smooth steps from the damping $\Gamma$. As the
field increases, the distribution shifts to higher wave-vector as predicted by the Boltzmann
theory. With higher field ($\Omega>\Gamma$), the distribution
develops strong nonlinear effect with increasing effective temperature.
}
\label{fig:nkt}
\end{figure}

To interpret the $k$-occupation number, we should study the quantities
with respect to the physically meaningful gauge-invariant (mechanical) wave-vector
$k_m=k+\Omega t$. The occupation number can be easily evaluated by replacing $k+\Omega t$
by $k_m$ in Eq.~(\ref{eq:nk}), as shown in FIG.~\ref{fig:nkt} for the
damping at $\Gamma=0.1$. As the field $\Omega$ increases, the
$k$-occupation number to the Fermi-Dirac distribution shifted towards
the field direction. Despite the lack of momentum scattering in the
system, the picture of displaced Fermi sea remains valid for small
field. The fermion thermostats acting as particle reservoirs seem to
dephase the Peierls factor when an electron is absorbed in the
reservoir, hence leading to the similar effect as the momentum
scattering. 
In appendix \ref{sec:nkt}, it has been shown analytically that the shift of the
wave-vector at small field is
\begin{equation}
\delta k = \frac{\Omega}{\Gamma}\propto E\tau,
\end{equation}
as expected in the Boltzmann transport picture. The momentum shift
$\delta k$, in the low-field limit, corresponds to the drift velocity
which is proportional to the electric field $E$ and the lifetime
$\tau(\sim\Gamma^{-1})$
of the transport electron given by the reservoir. As the field increases
the shift of Fermi surface deviates from the linear relation. As the
field is further increased ($\Omega\gg\Gamma$), the distribution significantly 
deviates from the sharp low temperature distribution and all $k_m$
gradually become equally occupied.

Another gauge-invariant quantities are the local variables. For instance,
by taking the $k$-summation of Eq.~(\ref{eq:nk}), one obtains the local
electron density. Due to the term $k+\Omega$ in the expression, the
average over $k\in[-\pi/a,\pi/a]$ is equivalent to the time-average over
$t\in[0,2\pi/\Omega]$, i.e., the local density becomes time-independent
for large time limit. Specifically, the $k$-summation requires $m=n$ and
we have
\begin{equation}
\bar{n}_{\rm local}(t)=
\frac12\sum_{m=-\infty}^{\infty}\left[
J_m\left(\frac{2\gamma}{\Omega}\right)
\right]^2=\frac12.\nonumber
\end{equation}

\begin{figure}
\includegraphics[width=\linewidth]{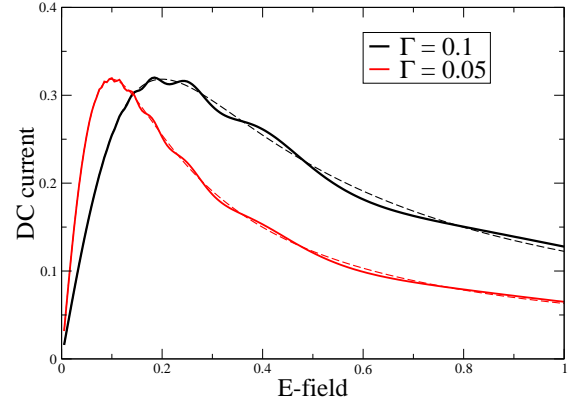}
\caption{(color online) DC current as a function of the damping $\Gamma$ and electric
field $\Omega=eEa$. For small field, the current has a linear dependence on
the $E$-field showing an Ohm's law-like behavior. As $E$ increases, the
Bloch oscillation behavior takes over and the DC current decreases. The
dashed lines are the simplified expression, Eq.~(\ref{eq:drude}).
}
\label{fig:omega}
\end{figure}

\begin{figure}
\includegraphics[width=\linewidth]{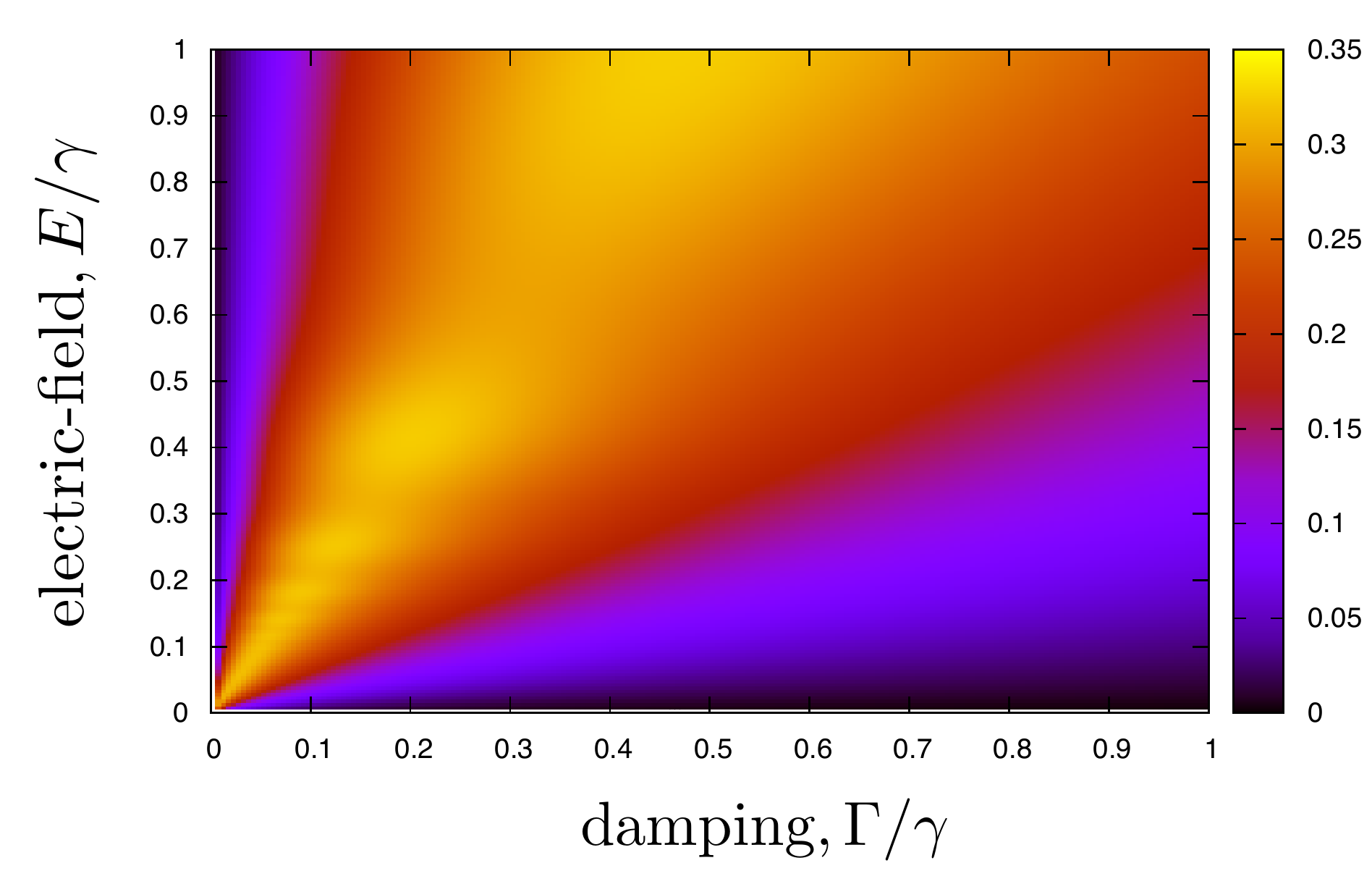}
\caption{(Color online) Contour plot of DC current as a function of damping and
electric field.
}
\label{fig4}
\end{figure}

Now we turn to the calculation of electric current,
\begin{equation}
J_k(t)=\frac{\partial\epsilon_d(k+\Omega t)}{\partial k}n_k(t)
=2\gamma\sin(k+\Omega t)n_k(t).
\end{equation}
Due to the sine-function, the DC current has contributions only from $m-n=\pm
1$ in Eq.~(\ref{eq:nk}). After some manipulations, we have
\begin{eqnarray}
\bar{J}_k & = &
\frac{2\gamma\Gamma}{\pi(\Omega^2+4\Gamma^2)}\sum_{m}J_m\left(\frac{2\gamma}{\Omega}\right)
J_{m-1}\left(\frac{2\gamma}{\Omega}\right)\times\nonumber\\
&&\left[\Gamma\log\frac{m^2\Omega^2+\Gamma^2}{(m-1)^2\Omega^2+\Gamma^2}
+\Omega\chi_{m,m-1}\right].
\label{eq:jk}
\end{eqnarray}
As in the case for $n_k(t)$, the DC limit of $J_k(t)$ becomes
independent of $k$. The total current is shown in Figs.~\ref{fig:omega} and
\ref{fig4}. Similar plot has been obtained in the interacting model
from numerical calculation of Hubbard model connected to fermion
bath~\cite{amaricci}.

It is instructive to simplify the above expression in the limit of
$\Omega,\Gamma\ll\gamma$ where
the DC current is reduced to the expression
\begin{equation}
\bar{J}\approx\frac{4\gamma\Gamma\Omega}{\pi(\Omega^2+4\Gamma^2)}.
\label{eq:drude}
\end{equation}
Detailed derivation is provided in Appendix. This approximate expression is shown
as dashed lines in Fig.~\ref{fig:omega}. Despite that the formula was
derived for $\Omega,\Gamma\ll\gamma$, it shows remarkable accuracy to the DC
current for a wide range of $\Gamma$ and $E$. 

It is also interesting to note that a similar formula has been
derived for a super-lattice system with Ohmic scattering within the
semi-classical Boltzmann transport equation~\cite{lebwohl}. Although the
current has the same dependence on the damping and the electric field,
it should be emphasized that the two models have quite different scattering
mechanism where in the Boltzmann approach~\cite{lebwohl} the momentum relaxation is
explicitly built-in while in our case the lattice wave-vector scattering
does not happen and a very different Fermi surface structure results.
In the low-field limit, the current~(\ref{eq:drude}) recovers the form of
the Drude conductivity per electron,
\begin{equation}
\bar{J}\approx\frac{\gamma\Omega}{\pi\Gamma}\propto \frac{E\tau}{m^*},
\end{equation}
with $\gamma\sim 1/m^*$ and $\Gamma\sim 1/\tau$.

\section{Conclusions}

Calculations on electron transport with fermion thermostats confirm
salient features of numerical results, and map the model to the
Boltzmann transport picture, such as Fermi surface shift in the
Brillouin zone by the drift velocity and the Ohmic-like limit of
electric current. In particular, the electric current,
Eq.~(\ref{eq:drude}), recovered the semi-classical transport result even
without any momentum scattering. Explicit and exact calculations clarify
the steady-state nature of the model which might have different
scattering processes from the realistic solid-state transport systems.
Nevertheless, its phenomenological similarity to the conventional
semi-classical pictures has been established. The findings lead us to
conclude that the fermion bath model, despite its drastic
simplifications in the one-particle coupling and the lack of momentum
scattering, can be considered as a rudimentary and minimal dissipation
mechanism which will be invaluable in further modeling strong correlation
physics through dynamical mean-field theory.

\section{Acknowledgements}
Author is grateful for helpful discussions with Kwon Park and Woo-Ram
Lee. This work has been supported by the National Science
Foundation with the Grant number DMR-0907150. Author also thanks the
Asia-Pacific Center for Theoretical Physics at Pohang, Korea, where
part of the work has been completed.

\appendix

\section{Momentum occupation number at small field}

\label{sec:nkt}

In Eq.~(\ref{eq:nkt}), we rewrite the expression by the gauge-invariant
wave-vector $k_m=k+\Omega t$ and the $\omega$-integral first to
obtain
\begin{equation}
\frac{i\Gamma}{\pi}\int_{-\infty}^0ds\int_{-\infty}^0ds'
\frac{e^{\Gamma(s+s')-i(2\gamma/\Omega)[\sin(k_m+\Omega s)-
\sin(k_m+\Omega s')]}}{s-s'+i\eta}.
\end{equation}
Since $(s-s'+i\eta)^{-1}={\cal P}(s-s')^{-1}-i\pi\delta(s-s')$ with the
principal part evaluation ${\cal P}$, the $\delta$-function part
yield a simple contribution of $\frac12$. Redefining the times by the
average time $T=\frac12(s+s')$ and the relative time $t_r=s-s'$, we can
express Eq.~(\ref{eq:nkt}) as
\begin{equation}
\frac12+\frac{i\Gamma}{\pi}\int_{-\infty}^0dT
\int_{-2|T|}^{2|T|}dt_r{\cal P}\frac{
e^{2\Gamma T-i(4\gamma/\Omega)\cos(k_m+\Omega T)\sin(\Omega
t_r/2)}}{t_r}.
\end{equation}
We look at the mechanical wave-vector slightly away from $\pm \pi/2$ and
set $k_m=\pi/2+\delta k$. We change variables as $y=-\Omega T$,
$x=\Omega t_r$. Then we have the integral as
\begin{equation}
\frac{i\Gamma}{\pi\Omega}\int_0^{\infty}dy
\int_{-2y}^{2y}dx{\cal P}\frac{
e^{-2(\Gamma/\Omega)y+i(4\gamma/\Omega)\sin(\delta k-y)\sin(x/2)
}}{x}.
\end{equation}
For small field $\Omega\ll\Gamma$, the integral has main contribution
from $|x|,y\leq \Omega/\Gamma\ll 1$. Then the integral can be approximately
evaluated for $|\delta k|\leq \Omega/\Gamma$ as
\begin{equation}
n(k_m)\approx\frac12+\frac{2\gamma\Omega}{\pi\Gamma^2}\left(
1-\frac{\Gamma}{\Omega}\delta k
\right).
\end{equation}

\section{Derivation of current at small field}

For both
$\Gamma,\Omega\ll\gamma$, we expand Eq.~(\ref{eq:jk}) to the
leading order of $m$ as,
\begin{eqnarray}
\bar{J}_k & \approx &
\frac{2\gamma\Gamma}{\pi(\Omega^2+4\Gamma^2)}\sum_{m}J_m\left(\frac{2\gamma}{\Omega}\right)
J_{m-1}\left(\frac{2\gamma}{\Omega}\right)\times\nonumber\\
&&\left[\frac{2m\Gamma\Omega^2}{m^2\Omega^2+\Gamma^2}
+2\Omega\tan^{-1}\frac{m\Omega}{\Gamma}\right].
\end{eqnarray}
Rearranging the summation and using the identity
$x[J_{m-1}(x)+J_{m+1}(x)]=2mJ_m(x)$, we write
\begin{eqnarray}
\bar{J}_k & \approx &
\frac{2\gamma\Gamma}{\pi(\Omega^2+4\Gamma^2)}\sum_{m}J_m
\left[J_{m-1}+J_{m+1}\right]\times\nonumber\\
&&\left[\frac{m\Gamma\Omega^2}{m^2\Omega^2+\Gamma^2}
+\Omega\tan^{-1}\frac{m\Omega}{\Gamma}\right]\\
&=&
\frac{2\Gamma\Omega}{\pi(\Omega^2+4\Gamma^2)}\sum_{m}m\Omega J_m^2
\left(\frac{m\Gamma\Omega}{m^2\Omega^2+\Gamma^2}
+\tan^{-1}\frac{m\Omega}{\Gamma}\right).
\nonumber
\end{eqnarray}
For $\Omega\ll\gamma$, we define $x=m\Omega$ in the regime
$m=x/\Omega\gg 1$, the summation becomes
\begin{equation}
\int_{-\infty}^\infty x
J_{\frac{x}{\Omega}}(\tfrac{2\gamma}{\Omega})^2
\left(\frac{x\Gamma}{x^2+\Gamma^2}
+\tan^{-1}\frac{x}{\Gamma}\right)\frac{dx}{\Omega}.
\nonumber
\end{equation}
Using the asymptotic expression~\cite{gradshteyn} for
$x/\Omega,\gamma/\Omega\to\infty$,
\begin{equation}
\left[J_\frac{x}{\Omega}(\tfrac{2\gamma}{\Omega})\right]^2\sim
\left\{\begin{array}{ll}
\frac{\Omega/2\gamma}{\pi\sqrt{1-(x/2\gamma)^2}} & (|x|<2\gamma)\\
0 & (|x|>2\gamma)
\end{array}\right.,
\nonumber
\end{equation}
the integral simplifies to
\begin{equation}
\int_{-2\gamma}^{2\gamma}
\frac{1}{\pi\sqrt{4\gamma^2-x^2}}
\left(\frac{x^2\Gamma}{x^2+\Gamma^2}
+x\tan^{-1}\frac{x}{\Gamma}\right)dx.
\nonumber
\end{equation}
In the limit $\Gamma\ll\gamma$, the second term in the parenthesis
dominates and we have
\begin{equation}
\bar{J}\approx\frac{2\Gamma\Omega}{\pi(\Omega^2+4\Gamma^2)}
\int_0^{2\gamma}\frac{xdx}{\sqrt{4\gamma^2-x^2}}
=\frac{4\gamma\Gamma\Omega}{\pi(\Omega^2+4\Gamma^2)}.
\end{equation}


\begin{thebibliography}{*}

\bibitem{kadanoff} Leo P. Kadanoff and Gordon Baym, \textit{Quantum
Statistical Mechanics}, Westview Press (1994).

\bibitem{mahan} G. D. Mahan, \textit{Many-Particle Physics} 3rd Ed.,
Chap. 8, Kluwer Academic (2000).

\bibitem{rosch} A. Rosch, J. Paaske, J. Kroha, and P. W\"olfle,
Phys. Rev. Lett. {\bf 90}, 076804 (2003).

\bibitem{schoeller} Herbert Schoeller and Gerd Scon, Phys. Rev.
B {\bf 50}, 18436 (1994).

\bibitem{mehta} P. Mehta and N. Andrei, Phys. Rev. Lett. {\bf 96},
216802 (2006).

\bibitem{werner} P. Werner, T. Oka, and A.J. Millis, Phys. Rev. B
{\bf 79}, 035320 (2009).

\bibitem{feiguin} F. Heidrich-Meisner, A.E. Feiguin, and E. Dagotto,
Phys. Rev. B {\bf 79}, 235336 (2009).

\bibitem{schiro} Marco Schiro and Michele Fabrizio, Phys. Rev. B
{\bf 79}, 153302 (2009).

\bibitem{boulat} E. Boulat, H. Saleur, and P. Schmitteckert, Phys.
Rev. Lett. {\bf 101}, 140601 (2008).

\bibitem{prl07} J. E. Han and R. J. Heary, Phys. Rev. Lett. {\bf 99},
236808 (2007).

\bibitem{anders} F. B. Anders, Phys. Rev. Lett. {\bf 101}, 066804 (2008).

\bibitem{turkowski} V. Turkowski and J. K. Freericks, Phys. Rev. B {\bf
71}, 085104 (2005).

\bibitem{freericks} J. K. Freericks, Phys. Rev. B {\bf 77}, 075109 (2008).

\bibitem{eckstein} Martin Eckstein, Takashi Oka, and Philipp Werner,
Phys. Rev. Lett. {\bf 105}, 146404 (2010).

\bibitem{aoki} Naoto Tsuji, Takashi Oka, and Hideo Aoki, Phys. Rev. B
{\bf 78}, 235124 (2008); Naoto Tsuji, Takashi Oka, and Hideo Aoki, 
Phys. Rev. Lett. {\bf 103}, 047403 (2009).

\bibitem{amaricci} A. Amaricci, C. Weber, M. Capone, and G. Kotliar,
Phys. Rev. B {\bf 86}, 085110 (2012).

\bibitem{aron} Camille Aron, Gabriel Kotliar, and Cedric Weber, Phys.
Rev. Lett. {\bf 108}, 086401 (2012).

\bibitem{vidmar} M. Mierzejewski, L. Vidmar, J. Bonca, and P. Prelovsek,
Phys. Rev. Lett. {\bf 106}, 196401 (2011); L. Vidmar, J. Bonca, T.
Tohyama, and S. Maekawa, Phys. Rev. Lett. {\bf 107}, 246404 (2011).

\bibitem{jauho} Antti-Pekka Jauho, Ned S. Wingreen and Yigal Meir, Phys.
Rev. B {\bf 50}, 5528 (1994).

\bibitem{blandin} A. Blandin, A. Nourtier, D. W. Hone, J. Phys.
(Paris) {\bf 37}, 369 (1976).

\bibitem{lebwohl} Paul A. Lebwohl and Raphael Tsu, J. Appl. Phys. {\bf
41}, 2664 (1970).

\bibitem{gradshteyn} I. S. Gradshteyn and I. M. Rhizyk, \textit{Table of
Integrals, Series, and Products}, formulae 8.452 and 8.453, 7th Ed.
Elsevier (2007).


\end{thebibliography}
\end{document}